\documentclass[12pt]{article}
\usepackage{epsf}
\title{ {\bf
The $b\rightarrow s g g$ decay in the general two Higgs doublet model}}
\author{\vspace{1cm}\\
        {\bf E. O. Iltan}
        \thanks{E-mail address:
        eiltan@heraklit.physics.metu.edu.tr}
 \\
        Physics Department, Middle East Technical University \\
        Ankara, Turkey\\}

\date{}

\begin{document}
\setlength{\baselineskip}{24pt}
\maketitle
\setlength{\baselineskip}{7mm}
\begin{abstract}
We study the decay width of the inclusive process 
$b\rightarrow s g g$ in the two Higgs doublet model with three level
flavor changing neutral currents (model III). We analyse the dependencies 
of the differential decay width to the $s$- quark energy $E_s$ and  model 
III parameters, charged Higgs mass $m_{H^{\pm}}$ and Yukawa coupling 
$\bar{\xi}_{N,bb}^D$. We observe that there exist a considerable enhancement 
in the decay width for the relevant process. This enhancement can be reduced 
by choosing $C_7^{eff}$ as negative and increasing the lower bound of 
$m_{H^{\pm}}$ to the large values, such as  $800\, GeV$ . This is an 
interesting result which gives an idea on the mass $m_{H^{\pm}}$ and sign 
of $C_7^{eff}$.
\end{abstract} 
\thispagestyle{empty}
\newpage
\setcounter{page}{1}
\section{Introduction}
The B-meson system is interesting and rich phenomenologically, providing a
comprehensive  information on the theoretical models. 
With the forthcoming experiments at SLAC, KEK
B-factories, HERA-B and possible future accelerators, the large number of
events can take place and various branching ratios of events, CP-violating
asymmetries, polarization effects, etc... can be measured \cite{Bar,Ellis1}. 
This will lead to test the models underconsideration and to determine 
corresponding free parameters.

Loop induced processes are sensitive to the parameters of the models used.
Therefore, they open a window for the determination 
of these parameters and investigation of new models. Among these type of 
decays, inclusive $b\rightarrow s g$ reached a great interest since it is 
theoretically clean and sensitive to new physics beyond SM, like two Higgs 
doublet model (2HDM)\cite{Glashow}, minimal supersymmetric Standard model 
(MSSM) \cite{Misiak1,Abel}, etc... . The Branching ratio $Br$ of 
$b\rightarrow s g$ decay in the SM is 
$Br(b\rightarrow s g)\sim 0.2 \%$ for on-shell gluon \cite{Gao} and  
the enhancement of this ratio brings an advantage \cite{Weinberg1} 
to decrease the averaged cham multiplicity  $\eta_c$ \cite{Grizad} 
and to increase kaon yields \cite{Kagan}. This enhancement can be 
obtained by including the QCD corrections or looking for new models 
beyond the SM. In the literature, there 
are number of theoretical calculations on the $Br$ of the corresponding 
process beyond the SM. In \cite{Chao1,Kagan2}, $Br\,(b\rightarrow s g)$ was 
calculated in the 2HDM  (Model I and II)  for $m_{H^{\pm}}\sim 200\, GeV$
and $tan\,\beta\sim 5$ and it was found that there was an enhancement less 
than one order. This decay was studied in the supersymmetric
models \cite{Berto} and further, the $Br$ was calculated in the framework of 
the model III \cite{Zhen}, resulting with the enhancement 
at least one order compared to the SM one. This make it possible to describe 
the results coming from experiments \cite{Kagan3}. 

In the case of time-like gluon, namely $b\rightarrow s g^*$ decay, $Br$ should be 
consistent with the CLEO data \cite{CLEO1}
\begin{eqnarray} 
Br\,(b\rightarrow s g^*) < 6.8\,\% 
\label{Br1}
\end{eqnarray}   
and in \cite{Zhen}, it was showed that the model III enhancement was not 
contradict with this data for light-like gluon case. 

As a further process, $g^*$  can decay into quark-quark $\bar{q}q$ or 
gluon-gluon (gg) pairs. Inclusive three body decay $b\rightarrow s g g$  
is another interesting one which is studied in the literature extensively 
\cite{Hou1,Simma,Zhen2}. It  becomes not only from the chain process 
$b\rightarrow s g^*$ followed by  
$g^*\rightarrow g g$ but also from the emission of on-shell gluons from the 
quark lines to obey gauge invariance. In \cite{Simma}, the complete 
calculation was done in the SM and $Br$ ratio was found at the order 
of $10^{-3}$. In \cite{Zhen,Zhen2} the additional contribution of gluon 
penguins in the Model III was estimated as negligible.         

This work is devoted to the study of the complete calculation for 
$b\rightarrow s g g$ decay in the model III and we find that the decay 
width ($\Gamma$) is strongly sensitive to the charged Higgs mass 
$m_{H^{\pm}}$. Therefore, it is possible to get a considerable enhancement 
in $\Gamma$ even 2 orders larger compared to the SM case.

The paper is organized as follows:
In Section 2, we give a brief summary for the model III. Further, we 
calculate the matrix element and decay width of the inclusive  
$b\rightarrow s g g $ decay in the framework of the model III. Section 3 
is devoted to discussion and our conclusions. In Appendix, we present 
the form factors appearing in the SM.
\section{The inclusive process $b\rightarrow s g g$ in the framework of the 
model III} 
The Yukawa interaction in the model III  can be defined as 
\begin{eqnarray}
{\cal{L}}_{Y}=\eta^{U}_{ij} \bar{Q}_{i L} \tilde{\phi_{1}} U_{j R}+
\eta^{D}_{ij} \bar{Q}_{i L} \phi_{1} D_{j R}+
\xi^{U}_{ij} \bar{Q}_{i L} \tilde{\phi_{2}} U_{j R}+
\xi^{D}_{ij} \bar{Q}_{i L} \phi_{2} D_{j R} + h.c. \,\,\, ,
\label{lagrangian}
\end{eqnarray}
where $L$ and $R$ denote chiral projections $L(R)=1/2(1\mp \gamma_5)$,
$\phi_{i}$ for $i=1,2$, are the two scalar doublets. The Yukawa matrices  
$\eta^{U,D}_{ij}$ and $\xi^{U,D}_{ij}$ have  in general complex entries.
With the choice of $\phi_{1}$ and $\phi_{2}$,
\begin{eqnarray}
\phi_{1}=\frac{1}{\sqrt{2}}\left[\left(\begin{array}{c c} 
0\\v+H^{0}\end{array}\right)\; + \left(\begin{array}{c c} 
\sqrt{2} \chi^{+}\\ i \chi^{0}\end{array}\right) \right]\, ; 
\phi_{2}=\frac{1}{\sqrt{2}}\left(\begin{array}{c c} 
\sqrt{2} H^{+}\\ H_1+i H_2 \end{array}\right) \,\, ,
\label{choice}
\end{eqnarray}
and the vacuum expectation values,  
\begin{eqnarray}
<\phi_{1}>=\frac{1}{\sqrt{2}}\left(\begin{array}{c c} 
0\\v\end{array}\right) \,  \, ; 
<\phi_{2}>=0 \,\, ,
\label{choice2}
\end{eqnarray}
the SM particles are collected in the first doublet and
particles due to new physics in the second one. The part of Yukawa 
interaction which is responsible for physics beyond the SM is the Flavor 
Changing (FC) interaction and can be written as 
\begin{eqnarray}
{\cal{L}}_{Y,FC}=
\xi^{U}_{ij} \bar{Q}_{i L} \tilde{\phi_{2}} U_{j R}+
\xi^{D}_{ij} \bar{Q}_{i L} \phi_{2} D_{j R} + h.c. \,\, ,
\label{lagrangianFC}
\end{eqnarray}
where the couplings  $\xi^{U,D}$ for the FC charged interactions are
\begin{eqnarray}
\xi^{U}_{ch}&=& \xi_{N} \,\, V_{CKM} \nonumber \,\, ,\\
\xi^{D}_{ch}&=& V_{CKM} \,\, \xi_{N} \,\, ,
\label{ksi1} 
\end{eqnarray}
and $\xi^{U,D}_{N}$ is defined by the expression (more details see 
\cite{Soni})
\begin{eqnarray}
\xi^{U,D}_{N}=(V_L^{U,D})^{-1} \xi^{U,D} V_R^{U,D}\,\, .
\label{ksineut}
\end{eqnarray}
Note that the index "N" in $\xi^{U,D}_{N}$ denotes the word "neutral". 

Now we start with the decay amplitude of the decay $b\rightarrow s g g$
\begin{eqnarray}
M(b\rightarrow s gg)=i \frac{\alpha_s\, G_F}{\sqrt{2}\pi}
\epsilon_a^{\mu}(k_1)\epsilon_b^{\nu}(k_2) \bar{s}(p')T^{a\,b}_{\mu\nu}\,
b(p)\,\, ,
\label{Amp1}
\end{eqnarray}
where $\epsilon_a^{\mu}(k)$ are polarization vectors of the gluons with
color $a$ and momentum $k$. Using the same parametrization for 
$T^{a\,b}_{\mu\nu}$ as in \cite{Simma}, we have
\begin{eqnarray}
T^{a\,b}_{\mu\nu}=T_{\mu\nu}\frac{\lambda^b}{2}\,\frac{\lambda^a}{2}+
T^{E}_{\mu\nu}\,\frac{\lambda^a}{2}\,\frac{\lambda^b}{2}\,\, ,
\label{Tfunc}
\end{eqnarray}
and  $T^{E}_{\mu\nu}$ can be obtained by the replacements 
$k_1\leftrightarrow k_2$ and $\mu\leftrightarrow \nu$ in the function 
$T_{\mu\nu}$. Here $\frac{\lambda^a}{2}$ are the Gell-Mann matrices.
The functions $T_{\mu\nu}$ and $T^{E}_{\mu\nu}$, in
general, contain masses of internal quarks $u,c,t$ in the SM and also 
$d,s,b$ in the model III, since the process underconsideration takes place 
at least at one loop level. Therefore, at this stage, we take into account 
two different possibilities, 
\begin{itemize}
\item the mass of internal quark is heavy (namely, $t$-quark), 
\item the mass of internal quark is light (namely, $d,s,b,u,c$-quarks). 
\end{itemize}

In the heavy internal quark case, the terms $k^2_{external}/m_i^2$ and 
$k^2_{external}/m_i^2\, (m^2_W,\, m^2_{H^{\pm}})$ are neglected and 
the form factors are obtained as functions of 
$x_t=m_t^2/m_W^2$ and $y_t=m_t^2/m_H^2$ where $m_{H^{\pm}}$ is the mass
of charged Higgs boson in the model III. Neglecting $s$-quark mass,
$T_{\mu\nu}$ for the heavy internal quark is given by
\begin{eqnarray}
T^{heavy}_{\mu\nu}&=&
-i \, \lambda_t \, F^{2HDM}_2 \, \{ \big ( \frac{2 \,p'_{\nu}+
\gamma_{\nu} \not\!{k_2}}{2 \,p'.k_2}\,
\sigma_{\mu\alpha} k_1^{\alpha}+\sigma_{\nu\alpha} k_2^{\alpha}\,
\frac{2 p_{\mu}-\not\!{k_1} \gamma_{\mu}}{-2 p.k_1} \big )\nonumber \\ 
&+& 
\frac{1}{q^2} \big ( 2\, \sigma_{\alpha\beta} k_{1}^{\alpha}\, 
k_2^{\beta} g_{\mu\nu}+2\, \sigma_{\nu\alpha} k_{2\,\mu}\, q^{\alpha}-
2\, \sigma_{\mu\alpha} k_{1\,\nu}\, q^{\alpha}+ 
\sigma_{\mu\nu} q^2 \big ) \} \,m_b\, R\,\, .  
\label{Tfunch}
\end{eqnarray}
Here $q$ is the momentum transfer, $q=k_1+k_2$, $\lambda_t$ is the CKM 
matrix combination $\lambda_t=V_{tb} V^{*}_{ts}$ and $F^{2HDM}_2$ is the 
form factor 
\begin{eqnarray}
F^{2HDM}_2=F^{SM}_2\,(x_t)+F^{Beyond}_2\,(y_t)
\label{Ffunc}
\end{eqnarray}
where $F^{SM}_2\,(x_t)$ is the magnetic dipole form factor of $b\rightarrow
s g^*$ vertex (see Appendix). $F^{Beyond}_2\,(y_t)$ is the contribution 
coming from the charged Higgs boson in the model III:
\begin{eqnarray}
F^{Beyond}_2\,(y_t)&=&\frac{1}{m_{t}^2} \,
(\bar{\xi}^{* U}_{N,tt}+\bar{\xi}^{* U}_{N,tc}
\frac{V_{cs}^{*}}{V_{ts}^{*}}) \, (\bar{\xi}^{U}_{N,tt}+\bar{\xi}^{U}_{N,tc}
\frac{V_{cb}}{V_{tb}}) G_{1}(y_t) \nonumber\, \, , \\
&-&\frac{1}{m_t m_b} \, (\bar{\xi}^{* U}_{N,tt}+\bar{\xi}^{* U}_{N,tc}
\frac{V_{cs}^{*}}{V_{ts}^{*}}) \, (\bar{\xi}^{D}_{N,bb}+\bar{\xi}^{D}_{N,sb}
\frac{V_{ts}}{V_{tb}}) G_{2}(y_t)
\, \, , 
\label{Fbeyond1}
\end{eqnarray}
and 
\begin{eqnarray}
G_1\,(y_t)&=& \frac{y_t}{12\, (-1+y_t)^4} 
\big( (-1+y_t)\,(-2-5 \,y_t+y_t^2)+6\, y_t\,ln\,y_t) \big )\nonumber 
\,\, ,\\ 
G_2\,(y_t)&=& \frac{1}{2\, (-1+y_t)^4} \,( y_t\,(3-4\, y_t+y_t^2) + 
2 \,(-1+y_t)\, y_t\, ln\,y_t) \,\, .
\label{G1G2}
\end{eqnarray}
In eq. (\ref{Fbeyond1}) we used the redefinition
\begin{eqnarray}
\xi^{U,D}=\sqrt{\frac{4 G_{F}}{\sqrt{2}}} \,\, \bar{\xi}^{U,D}\,\, .
\label{ksidefn}
\end{eqnarray}
Note that we neglect the chiral partner of the form factor  
$F^{Beyond}_2\,(y_t)$  and the neutral Higgs boson effects  which should 
be very small due to the discussion given in \cite{alil2} 
(see also Discussion part).

If the internal quark is light ($u$ or $c$), the first additional 
contribution comes from $m_{i}^2/m_W^{2}$ and 
$m_{i}^2/m_{H^{\pm}}^{2}$ terms. In the approximation 
$m_{i}^2/m_{W\,(H^{\pm})}^{2}\rightarrow 0$, it is enough to
replace $F^{SM}_2\,(x_t)$ with $\, "-F_2(0)"\,$ since
$\lambda_{c}=-\lambda_t$ by unitarity, namely $\sum_{i=u,c,t}
\lambda_i=0$. There is no additional term coming from light quark 
for $F^{Beyond}_2\,(y_t)$, since $F^{Beyond}_2\,(0)$ almost vanishes.    
For light internal quark, the second additional contribution 
comes from $k_{external}^2/m_i^{2}$ term which can not be neglected as 
in the heavy internal quark case. This contribution was calculated in the 
literature \cite{Simma} and we give its explicit form in Appendix.  
Therefore, the resulting amplitude can be written as 
\begin{eqnarray}
T_{\mu\nu}=T^{heavy}_{\mu\nu}+T^{light}_{1\,\mu\nu}
+T^{light}_{2\,\mu\nu}\,\, ,
\label{Tfunctot1}
\end{eqnarray}
and $T^{heavy}_{\mu\nu}$ is given in eq. (\ref{Tfunch}), 
$T^{light}_{1\,\mu\nu}$ can be obtained from $T^{heavy}_{\mu\nu}$ 
with the replacement $F_2^{2HDM} \rightarrow -F_2^{SM}(0)$ and $
T^{light}_{2\,\mu\nu}$ is given in Appendix.

The decay amplitude for the process $b\rightarrow s g g$, 
$T^{a\,b}_{\mu\nu}$, can be parametrized by seperating color symmetric 
and antisymmetric parts \cite{Simma} as
\begin{eqnarray}
T_{\mu\nu}^{a\,b}=T_{\mu\nu}^{+} \{ \frac {\lambda^{b}}{2}, 
\frac {\lambda^{a}}{2} \}+ T_{\mu\nu}^{-}[\frac {\lambda^{b}}{2},
\frac {\lambda^{a}}{2}]\,\, ,
\label{Tfunc2}
\end{eqnarray}  
with
\begin{eqnarray}
T_{\mu\nu}^{+}&=&\frac{1}{2}(T_{\mu\nu}+T_{\mu\nu}^{E}) 
\nonumber\,\, ,\\
T_{\mu\nu}^{-}&=&\frac{1}{2}(T_{\mu\nu}-T_{\mu\nu}^{E})\,\, . 
\label{Tfunc3}
\end{eqnarray}

Finally we get the differential decay width of the process using the
expression
\begin{eqnarray}
\frac{d^2\,\Gamma}{dE_s\,dE_1}=\frac{1}{2\pi^3}\frac{1}{8\, m_b}
|\bar{M}|^2\,\, ,
\label{diffCS}
\end{eqnarray}
where $E_s$ is the $s$-quark energy and $E_1$ is the energy of photon with
polarization $\epsilon_{\mu}^{a}(k_1)$. Here $\bar{M}$ is the average 
decay amplitude $\bar{M}=\frac{1}{2\,J+1}\,\frac{1}{N_c}\, M$ and 
$J=\frac{1}{2}$, $N_c=3$. 
Now, we divide the differential decay width into sectors as
follows:
\begin{itemize}
\item Symmetric sector, ($\Gamma^{Sym}$),
\item Antisymmetric sector, ($\Gamma^{Asym}$),
\end{itemize}
or 
\begin{itemize}
\item Right sector, ($\Gamma^{R}$),
\item Left sector, ($\Gamma^{L}$),
\item Left-rigth mixed sector, ($\Gamma^{LR}$).
\end{itemize}
Antisymmetric and symmetric sectors do not mix and they enter into
decay width as 
\begin{eqnarray}
\Gamma^{Sym\,(Asym)} \sim Tr(T_{\mu\nu}^{+\,(-)}\, (\not\!{p}+m_b))\,
\bar{T}_{\mu '\nu '}^{+\,(-)}\,\not\!{p'})\,P^{\mu\mu '}\,P^{\nu\nu '} 
\,\, , 
\end{eqnarray}
with the corresponding color factors 
$C_+=\frac{(N_c^2-1)(N_c^2-2)}{2\, N_c}$ and $C_-=\frac{N_c\,(N_c^2-1)}{2}$\, 
respectively. Here we choose the polarization sum of the on-shell gluons as
\begin{eqnarray}
P^{\mu\mu '}=-g^{\mu\mu '}+\frac{ k_1^{\mu}\,k_2^{\mu '}+
k_2^{\mu}\,k_1^{\mu '}}{k_1.k_2} \nonumber \,\, ,
\end{eqnarray}
and $\bar{T}_{\mu '\nu '}^{+\,(-)}= 
\gamma_0\,(T_{\mu '\nu '}^{+\,(-)})^{\dagger}\,\gamma_0$. Right sector 
contains form factors which are functions of $x_i=m_i^2/m_W^2$ and 
$y_i=m_i^2/m_H^2$ where $i=u,c,t$.  Left one have the form factors 
which are created by the nonvanishing $k_{external}^2/m_{light}^2$ terms. 
Left-right sector contains mixed terms and its contribution is negligible.

Since there are collinear divergences at the boundary of the kinematical 
region, we follow the procedure given in \cite{Simma}, namely taking a 
cutoff $c$ in the integration over phase space: 
\begin{eqnarray}
 \frac{-2\,E_s\,m_b+m_b^2+m_s^2}{2\, m_b} + c\, m_b \leq \, E_1\,\leq
\frac{-2\,E_s\,m_b+m_b^2}{2\,(2\,E_s- m_b)} \,\, ,
\label{cutoff}
\end{eqnarray}
with $c=0.01$.
Note that these limits are used in the integration over $E_1$ and to get
differential decay width $\frac{d\,\Gamma}{d\,E_s}$.
\section{Discussion}
There are many free parameters in the model III such as  Yukawa couplings, 
$\xi_{ij}^{U,D}$ where $i$, $j$ are flavor indices, masses of charged and 
neutral Higgs bosons. The procedure is to restrict these parameters using 
the experimental measurements. Since the contributions of the neutral Higgs 
bosons $h_0$ and $A_0$ to the Wilson coefficient $C_7^{eff}$ should not 
contradict with the CLEO measurement \cite{CLEO2}, 
\begin{eqnarray}
Br (B\rightarrow X_s\gamma)= (3.15\pm 0.35\pm 0.32)\, 10^{-4} \,\, ,
\label{br2}
\end{eqnarray}
the couplings $\bar{\xi}^{D}_{N,is}$($i=d,s,b)$ and 
$\bar{\xi}^{D}_{N,db}$ should be  negligible (see \cite{alil2} for 
details). In addition, the constraints \cite{alil1}, coming from the 
$\Delta F=2$ mixing, the $\rho$ parameter \cite{atwood}, 
and the measurement by CLEO Collaboration results in the following
restrictions:   $\bar{\xi}_{N,tc} << \bar{\xi}^{U}_{N,tt},
\,\,\bar{\xi}^{D}_{N,bb}$ and $\bar{\xi}^{D}_{N,ib} \sim 0\, , 
\bar{\xi}^{D}_{N,ij}\sim 0$, where the indices $i,j$ denote $d$ and $s$ 
quarks. Therefore, we can  neglect all the couplings except 
$\bar{\xi}^{U}_{N,tt}$ and $\bar{\xi}^{D}_{N,bb}$. This leads to the 
cancellation of the contributions coming from primed coefficient 
$F_2^{' Beyond}$ and from the neutral Higgs  bosons $h_0$ and $A_0$, 
having interactions which include the Yukawa vertices with the 
combinations of $\bar{\xi}^{D}_{N,sb}$ and $\bar{\xi}^{D}_{N,ss}$.
Finally, we only take into account the multiplication of Yukawa 
couplings, $\bar{\xi}^{U}_{N,tt}\,\bar{\xi}^{* D}_{N,bb}$ and  
$|\bar{\xi}^{U}_{N,tt}|^2$ in our expressions.

In this section, we study the $s$ quark energy $E_s$, Yukawa coupling
$\bar{\xi}_{N,bb}^D$ and charged Higgs mass $m_{H^\pm}$ dependencies of the 
differential decay width $\frac{d\,\Gamma}{d\,E_s}$ for the inclusive decay
$b\rightarrow s g g$. In our analysis, we restrict the parameters  $
\bar{\xi}^{U}_{N,tt}$, $\bar{\xi}^{D}_{N bb}$ using the constraint for 
$|C_7^{eff}|$, $0.257 \leq |C_7^{eff}| \leq 0.439$ \cite{CLEO2}, where 
the upper and lower limits were calculated in \cite{alil1}
following the procedure given in \cite{gudalil}. 
Throughout these calculations, we take the charged Higgs mass
$m_{H^{\pm}}=400\, GeV$, and we use the input values given in Table 
(\ref{input}).  
\begin{table}[h]
        \begin{center}
        \begin{tabular}{|l|l|}
        \hline
        \multicolumn{1}{|c|}{Parameter} & 
                \multicolumn{1}{|c|}{Value}     \\
        \hline \hline
        $m_c$                   & $1.4$ (GeV) \\
        $m_b$                   & $4.8$ (GeV) \\           
        $\lambda_t$            & 0.04 \\
        $m_{t}$             & $175$ (GeV) \\
        $m_{W}$             & $80.26$ (GeV) \\
        $m_{Z}$             & $91.19$ (GeV) \\
        $\Lambda_{QCD}$             & $0.214$ (GeV) \\
        $\alpha_{s}(m_Z)$             & $0.117$  \\
        $c$                  & $0.01$  \\
        \hline
        \end{tabular}
        \end{center}
\caption{The values of the input parameters used in the numerical
          calculations.}
\label{input}
\end{table}

\begin{figure}[htb]
\vskip -3.0truein
\centering
\epsfxsize=6.8in
\leavevmode\epsffile{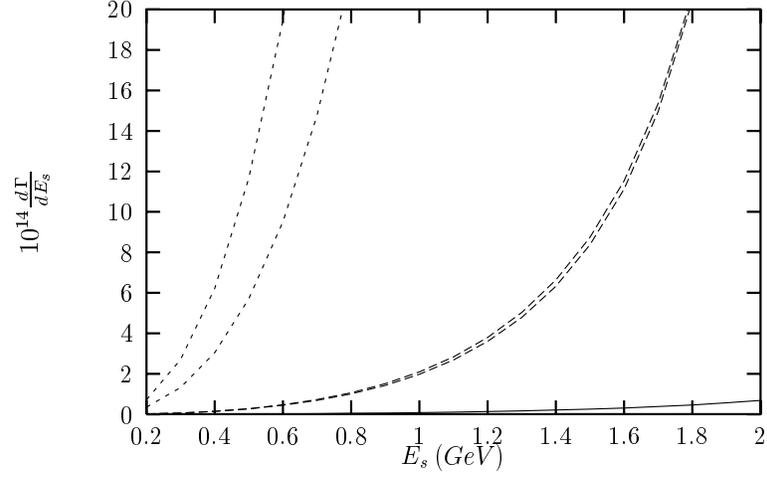}
\vskip -3.0truein
\caption[]{$\frac{d\,\Gamma}{d\,E_s}$ a function of 
$E_s$  for fixed $\bar{\xi}_{N,bb}^{D}=40\, m_b$ and  
$|r_{tb}|=|\frac{\bar{\xi}_{N,tt}^{U}}{\bar{\xi}_{N,bb}^{D}}| <1.$
Here $\frac{d\,\Gamma}{d\,E_s}$  is restricted in the region bounded 
by dotted (dashed) lines for $C_7^{eff} > 0$ ($C_7^{eff} < 0$). 
Solid line represents the SM contribution.}
\label{Tau2}
\end{figure}

In  Fig.~\ref{Tau2} we plot $\frac{d\,\Gamma}{d\,E_s}$ with respect to 
the $s$ quark energy $E_s$, for $\bar{\xi}_{N,bb}^{D}=40\, m_b$, 
and $|r_{tb}|=|\frac{\bar{\xi}_{N,tt}^{U}}{\bar{\xi}_{N,bb}^{D}}| <1.$
$\frac{d\,\Gamma}{d\,E_s}$  is restricted in the region bounded by dotted
(dashed) lines for $C_7^{eff} > 0$ ($C_7^{eff} < 0$). Solid line represents 
the SM contribution. There is a considerable enhancement in the differential
decay width especially for $C_7^{eff} > 0$ case. Besides, the allowed region 
becomes larger for $C_7^{eff} > 0$.

\begin{figure}[htb]
\vskip -3.0truein
\centering
\epsfxsize=6.8in
\leavevmode\epsffile{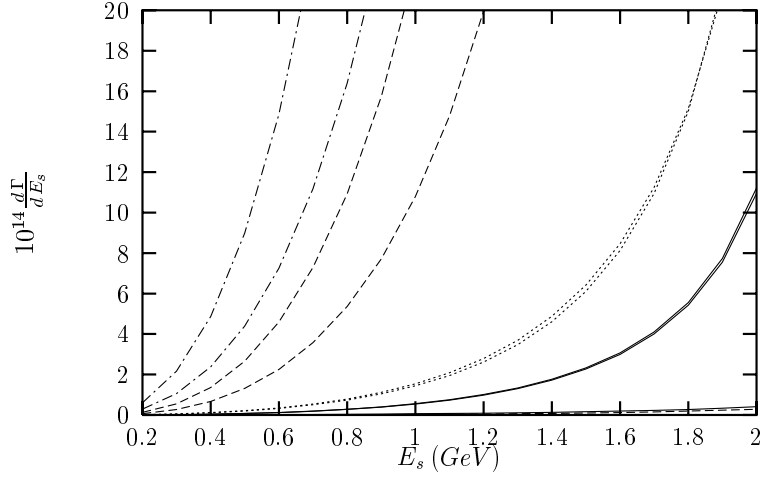}
\vskip -3.0truein
\caption[]{The color antisymmetric and symmetric part of 
$\frac{d\,\Gamma}{d\,E_s}$ as a function of 
$E_s$  for fixed $\bar{\xi}_{N,bb}^{D}=40\, m_b$,
$m_{H^{\pm}}=400\, GeV$ and $|r_{tb}|<1.$ 
Here the color antisymmetric part lies in the region bounded 
by dash-dotted (dotted) lines and  the color symmetric part by dashed  
(solid) lines, for $C_7^{eff} > 0$  ($C_7^{eff} < 0$). The SM contribution 
almost coincides with x-axis.}
\label{TauASySy}
\end{figure}

Fig.~\ref{TauASySy} is devoted to the $E_s$ dependence of color 
antisymmetric and symmetric part of $\frac{d\,\Gamma}{d\,E_s}$.
The color antisymmetric part lies in the region bounded by dash-dotted 
(dotted) lines and  the color symmetric part by dashed  (solid) lines, 
for $C_7^{eff} > 0$  ($C_7^{eff} < 0$). This figure shows that, for 
$C_7^{eff} > 0$, the contribution of the color antisymmetric part is 
greater than that of color symmetric one. This is true also  for 
$C_7^{eff} < 0$ case. However the contribution of the color symmetric 
part for $C_7^{eff} > 0$ exceeds  that of the color antisymmetric one for 
$C_7^{eff} < 0$. The allowed region becomes narrower for $C_7^{eff} < 0$ 
(see dotted and solid lines). Note that the contributions due to the SM 
is presented by the solid and dashed lines which almost coincide with the 
x-axis. 
 
\begin{figure}[htb]
\vskip -3.0truein
\centering
\epsfxsize=6.8in
\leavevmode\epsffile{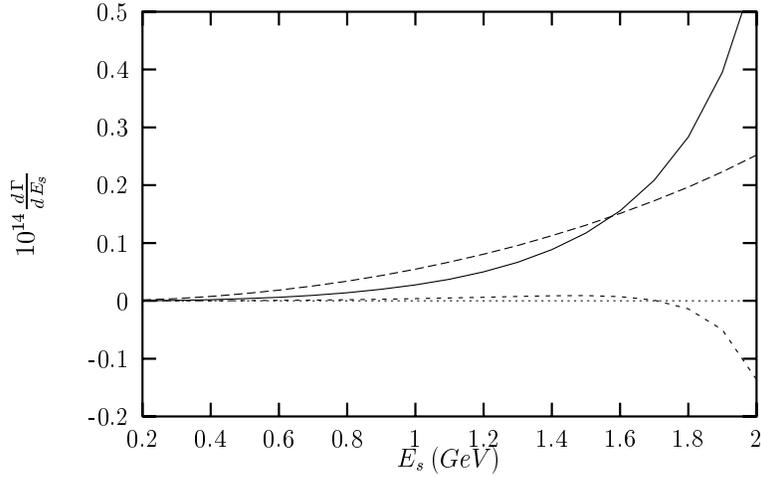}
\vskip -3.0truein
\caption[]{Right,  left and left-right mixed parts of 
$\frac{d\,\Gamma}{d\,E_s}$ a function of 
$E_s$  for fixed $\bar{\xi}_{N,bb}^{D}=40\, m_b$, $m_{H^{\pm}}=400\, GeV$ 
and $|r_{tb}|<1.$ Here solid line represents right, dashed line left 
and dotted line left-right  contributions.} 
\label{TauRiLeSM}
\end{figure}

Fig. \ref{TauRiLeSM} shows the $E_s$ dependence of right, 
left and left-right mixed parts of $\frac{d\,\Gamma}{d\,E_s}$ in the SM. 
Solid line represents right, dashed line left and dotted line left-right  
contributions. The left one exceeds the right one up to almost 
$E_s=1.6\, GeV$ since the $k^2_{external}/m^2_{light}$ 
contribution, responsible for left part, is comparable with the heavy 
internal quark, namely $m_t$, contribution. Left-right mixed part is very 
small and has also negative values. For the model III we have only right 
additional contributions since there is no left part beyond the SM in 
our approximation.  

\begin{figure}[htb]
\vskip -3.0truein
\centering
\epsfxsize=6.8in
\leavevmode\epsffile{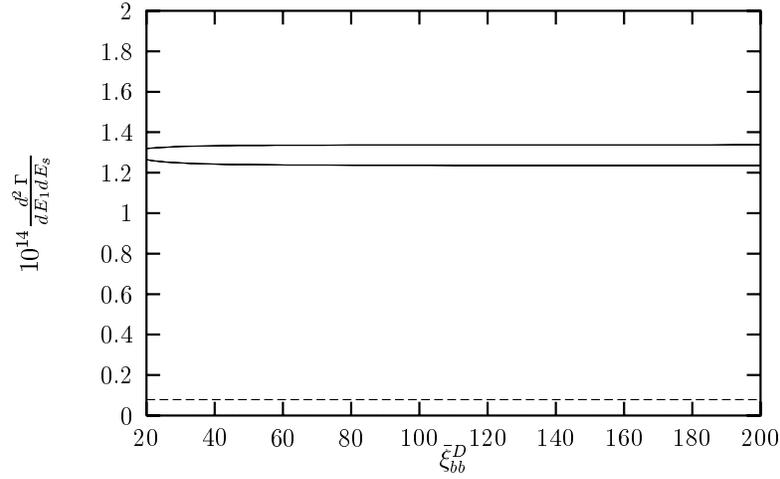}
\vskip -3.0truein
\caption[]{ $\frac{d\,\Gamma}{d\,E_s}$ as a function of $\bar{\xi}_{N,bb}^D$ 
for fixed $\bar{\xi}_{N,bb}^{D}=40\, m_b$, $m_{H^{\pm}}=400\, GeV$, 
$|r_{tb}|<1$, $C_7^{eff} < 0$, $E_1=2\,GeV$  and $E_s=1\,GeV$. 
Here $\frac{d\,\Gamma}{d\,E_s}$ lies in the region bounded by solid lines.
Dashed line represents the SM contribution.}
\label{TauDbbE12Es1a}
\end{figure}

\begin{figure}[htb]
\vskip -3.0truein
\centering
\epsfxsize=6.8in
\leavevmode\epsffile{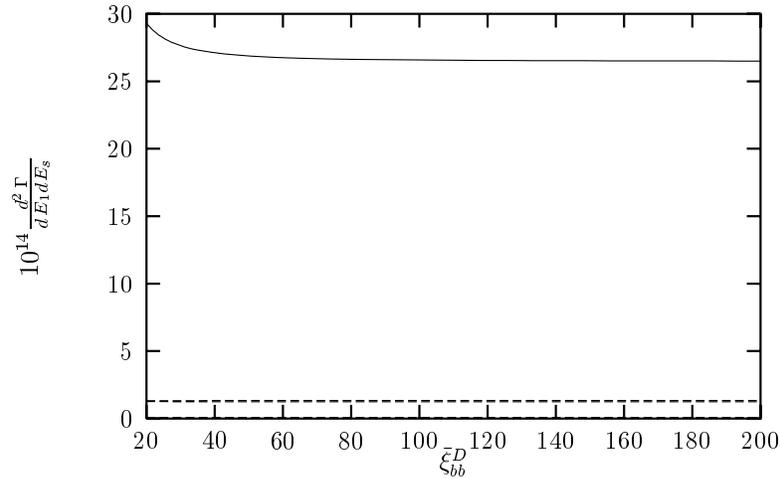}
\vskip -3.0truein
\caption[]{The same as Fig. \ref{TauDbbE12Es1b}, but for $C_7^{eff} > 0$.
Dashed line represents the contribution for $C_7^{eff} < 0$}
\label{TauDbbE12Es1b}
\end{figure}

In  Fig.~\ref{TauDbbE12Es1a} and Fig.~\ref{TauDbbE12Es1b}, we present 
the $\bar{\xi}_{N,bb}^D$ dependence of of $\frac{d\,\Gamma}{d\,E_s}$ for 
fixed values of $E_1=2\,GeV$ and $E_s=1\,GeV$. It is seen that there is 
almost no dependence on the parameter $\bar{\xi}_{N,bb}^D$ especially for 
its large values.   

\begin{figure}[htb]
\vskip -3.0truein
\centering
\epsfxsize=6.8in
\leavevmode\epsffile{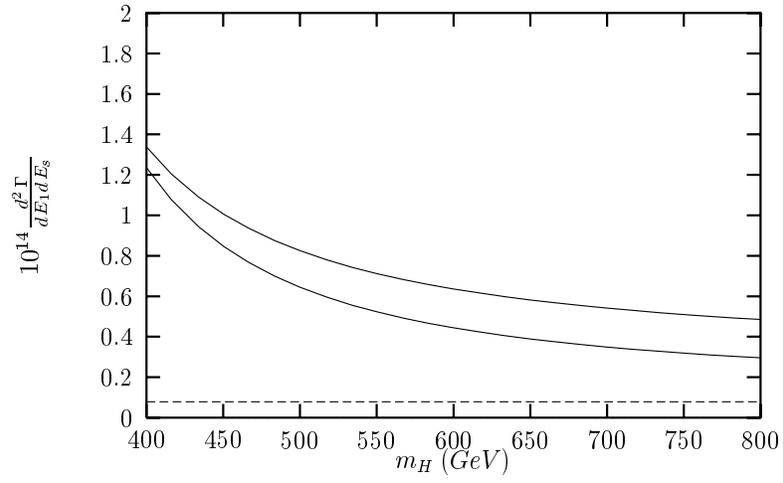}
\vskip -3.0truein
\caption[]{The same as Fig \ref{TauDbbE12Es1a}, but 
$\frac{d\,\Gamma}{d\,E_s}$ as a function of $m_{H^{\pm}}$ and for fixed 
$\bar{\xi}_{N,bb}^{D}=40\, m_b$.}
\label{TaumhE12Es1}
\end{figure}

For completeness, we also present $m_{H^{\pm}}$ dependence of  
$\frac{d\,\Gamma}{d\,E_s}$ for fixed values of 
$\bar{\xi}_{N,bb}^D=40\,m_b$, $E_1=2\,GeV$ and $E_s=1\,GeV$ for 
$C_7^{eff} < 0$ (Fig.~\ref{TaumhE12Es1}). Here the 
restricted region is bounded by solid lines. This figure shows that 
there is a strong dependence on the mass $m_{H^{\pm}}$. 

Now we would like to give some numerical results for our calculations. 
The total decay width for $b\rightarrow s X$ transition is 
\begin{eqnarray}
\Gamma_{tot}=(r\, |V_{ub}|^2\,+s \, |V_{cb}|^2)\Gamma_0 \,\, , 
\label{gamma}
\end{eqnarray}
where $\Gamma_0=\frac{m_b^5\, G_F^2}{192 \pi^3}$ and $r,s $ are 
QCD sensitive parameters \cite{Pashos}
\begin{eqnarray}
& & 6.46 \leq r \leq 7.55 \nonumber \,\, , \\
& & 2.38 \leq r \leq 2.92 \nonumber
\label{rs}
\end{eqnarray}
for $\alpha_s=0.2$. 
Eeasy calculation shows that the total decay width is 
$\Gamma_{tot}= 3.50\pm\,1.50\,\, 10^{-13}$. 

In our calculation, we obtain the decay width for the SM as 
$\Gamma_{SM}=5.1\,\, 10^{-15}\, GeV$. For $m_{H^{\pm}}=400\, GeV$ and 
$\bar{\xi}_{N,bb}^D=40\, m_b$, the model III result is 
four (three) orders larger for $C_7^{eff} > 0$ ($C_7^{eff} < 0$) 
compared to the SM result. This is a strong enhancement contradict 
with the total decay width given above. This forces us to choose the 
sign of  $C_7^{eff}$ as negative  ($C_7^{eff} < 0$)  and also to take 
large values of charged Higgs mass, 
$m_{H^{\pm}}$. For $m_{H^{\pm}}=800\, GeV$,  $\bar{\xi}_{N,bb}^D=40\, m_b$ 
and  $C_7^{eff} < 0$ we get:
\begin{eqnarray}
& & 8.58\, 10^{-14}\, GeV \leq \Gamma \leq 1.5\,10^{-13}
\, GeV \nonumber \,\, , \\ 
& & 3.09\, 10^{-14}\, GeV \leq \Gamma^{Sym}\leq 2.70\,10^{-14}
\, GeV \nonumber \,\, , \\  
& & 5.49\, 10^{-14}\, GeV \leq \Gamma^{ASym} \leq 1.23\,10^{-13}
\, GeV \,\, , \\ 
& & 8.08\, 10^{-14}\, GeV \leq \Gamma^{R} \leq 1.45\,10^{-13}
\, GeV \nonumber \,\, , \\ 
& & 4.93\, 10^{-15}\, GeV \leq \Gamma^{L} \leq 4.93\,10^{-15}
\, GeV \nonumber \,\, . 
\end{eqnarray}
   
In conclusion, we get a considerable enhancement in the decay width of the
process $b\rightarrow s g g $ in the model III. 
The enhancement can be suppressed by choosing  $C_7^{eff} < 0$ and 
increasing lower bound of charged Higgs mass, $m_{H^{\pm}}$. Further, 
the decay width of the process under consideration is not sensitive to the 
parameter $\bar{\xi}_{N,bb}^D$. 
\newpage
{\bf \LARGE {Appendix}} \\
\begin{appendix}
\section{The form factors in the SM for $b\rightarrow s g^*$ decay}
Here we present the magnetic dipole form factor $F^{SM}_2\,(x_t)$ and the
additional form factors due to the non-vanishing
$k_{external}^2/m_{light}^2$ terms. (for details see \cite{Simma}).
The vertex function for $b\rightarrow s g^*$ decay with on-shell quarks 
can be written as 
\begin{eqnarray}
\Gamma_{\mu}(p,p',q)=F_1\,(x_t)\, (q^2\gamma_{\mu}-q_{\mu} \not\!{q})\, 
L- F_2\,(x_t)\, i\, \sigma_{\mu\nu}\, q^{\nu}\, (m_b\, R + m_s \, L)\,\,,
\label{vertex1}
\end{eqnarray}
where $p$, $p'$ and $q$ are four-momentum of $b$-quark, $s$-quark and 
gluon respectively. The magnetic dipole form factor $F^{SM}_2\,(x_t)$  
in the SM is  
\begin{eqnarray}
F_2^{SM}(x_t)=\frac {-8+38 \,x_{t}-39 \,x^{2}_{t}+14\, x^{3}_{t}-5\,
x^{4}_{t}+18\, x^{2}_{t} \,ln\, x_{t}} {12\,(-1+x_t)^4}\,\, ,
\label{F2SM}
\end{eqnarray}
and $x_t=m_t^2/m_W^2$. The non-vanishing $k_{external}^2/m_{light}^2$ 
terms for light quarks bring new additional contributions, 
$\Delta\, F_1$, $\Delta\, i_2$, and $\Delta\, i_5$ (See \cite{Simma} for
details):
\begin{eqnarray}
\Delta F_{1}&=&-{\frac{2}{9}}-{\frac{4}{3}} \frac{Q_{0}(z)}{z}-{\frac{2}{3}} 
Q_{0}(z)\nonumber \,\, , \\
\Delta i_{2}&=&-{\frac{5}{9}}-2 \frac{Q_{-}(z)}{z}+{\frac{8}{3}}
\frac{Q_{0}(z)}{z}-{\frac{2}{3}}Q_{0}(z) \nonumber\,\, ,\\
\Delta i_{5}&=&-1-2  \frac{Q_{-}(z)}{z} \,\, ,
\label{F1i2i5}
\end{eqnarray}
where
\begin{eqnarray}
Q_{0}(z)&=&-2-(u_{+}-u_{-})(ln \frac {u_{-}}{u_{+}}+i\pi)\nonumber\,\, , \\
Q_{-}(z)&=&\frac{1}{2}(ln \frac {u_{-}}{u_{+}}+i\pi)^2\,\, , 
\label{Q0mp} 
\end{eqnarray}
with
\begin{eqnarray}
u_{\pm}=\frac{1}{2}(1\pm\sqrt {1-\frac{4}{z}}) \,\, ,
\label{upm}      
\end{eqnarray}
and
\begin{eqnarray}
z=\frac{q^2}{m_i^2},\,\,\, i=u,c \,\, .
\label{zz}
\end{eqnarray}

Finally, the contributions due to the  non-vanishing 
$k_{external}^2/m_{light}^2$ terms are
\begin{eqnarray}
T^{light}_{2\,\mu\nu}&=&-\lambda_t\, \{(\Delta\,i_2-\Delta\, F_1) 
(\not\!{k_1}-\not\!{k_2})\, g_{\mu\nu}\, L+ \Delta\, i_5\, i\, 
\epsilon_{\alpha\mu\nu\beta} \gamma^{\beta} 
(k_1^{\alpha}-k_2^{\alpha})\,L\nonumber \\ &-&2 \Delta\,F_1\, 
(\gamma_{\nu}\,k_{2\,\mu}-\gamma_{\mu}\,k_{1\,\nu})\,L \}
\label{Tlight2}
\end{eqnarray}

\end{appendix}
\newpage

\end{document}